# Context-Oriented Programming: A Programming Paradigm for Autonomic Systems

GUIDO SALVANESCHI and CARLO GHEZZI and MATTEO PRADELLA, Politecnico di Milano

Dynamic software adaptability is one of the central features leveraged by autonomic computing. However, developing software that changes its behavior at run time adapting to the operational conditions is a challenging task. Several approaches have been proposed in the literature to attack this problem at different and complementary abstraction levels: software architecture, middleware, and programming level. We focus on the support that ad-hoc programming language constructs may provide to support dynamically adaptive behaviors. We introduce context-oriented programming languages and we present a framework that positions the supported paradigm in the MAPE-K autonomic loop. We discuss the advantages of using context-oriented programming languages instead of other mainstream approaches based on dynamic aspect-oriented programming languages. We present a preliminary case study that shows how the proposed programming style naturally fits dynamic adaptation requirements and we extensively evaluate the use of COP in this scenario. Finally, we discuss some known problems and outline a number of open research challenges.

## 1. INTRODUCTION

In the course of years, software systems complexity has been growing, increasing the required effort and the cost of management and maintenance. Autonomic computing (AC) [Kephart and Chess 2003] aims at designing and building systems that can manage themselves with reduced human intervention pursuing high-level administrator's goals. Since these goals must be achieved in an evolving environment, autonomic systems must adjust their activity while running. That is, they are required to be *self-adaptable*, i.e., capable of modifying their behavior depending on the decisions taken by a managing component.

The studies on adaptable systems benefit from the contributions from a wide range of disciplines, including control theory and artificial intelligence. A lot of research has been done in this direction also from the software engineering community, since the design, development and maintenance of adaptable systems is especially challenging.

The problem has been faced with a variety of solutions leveraging reconfigurable architectures [Oreizy et al. 2008], component-based design [McKinley et al. 2004], and middleware [Liu and Martonosi 2003]. Other approaches tackled the problem at the programming language level. Language-level approaches are of special interest because they push the adaptation down to the elementary components of software, allowing for extremely fine-grain adaptability. The mainstream approaches have been focusing on *aspect-oriented programming* (AOP) [Kiczales et al. 2001] to enforce separation of concerns and *dynamic aspect-oriented programming* (DAOP) [Popovici et al. 2003] to support run-time adaptation. In particular DAOP, since its introduction in the AC field by Greenwood and Blair [Greenwood and Blair 2003], became a reference paradigm for autonomic systems due to its support for dynamic software modifications. The main research contributions that apply DAOP to AC are reviewed later in Section 7.

Aspect-oriented techniques support an effective management of separate concerns, such as the monitoring functionalities, which constitute an essential AC feature. DAOP makes it easy to add autonomic capabilities to an existing system with minimal impact on the codebase. However, we will

---

This research has been funded by the European Community's IDEAS-ERC Programme, Project 227977 (SMSCom).
Author's addresses: G. Salvaneschi and C. Ghezzi and M. Pradella, DEEPSE Group, DEI, Politecnico di Milano, Piazza L. Da Vinci, 32 Milano, Italy. {salvaneschi,ghezzi,pradella}@elet.polimi.it

show that context-oriented programming (COP) allows for better design-for-dynamic-adaptation approaches than DAOP. Starting from the pioneering work of Costanza and Hirschfeld [Costanza and Hirschfeld 2005], COP emerged as a promising paradigm for developing applications whose behavior can automatically adapt to changes in the context in which the application is embedded and running. This is especially common in the field of ubiquitous and pervasive computing. COP provides specific language-level abstractions to define and activate behavioral variations that allow the application to dynamically adapt to changes in the execution context.

This paper proposes and motivates the adoption of the COP paradigm in the field of AC. For this purpose, we introduce a methodology that conceptually separates context provisioning from the execution of the adaptable software in the COP paradigm, mapping the first on the autonomic manager and the second on the managed element of the MAPE-K model. We argue that COP can be effectively used to implement autonomic systems and that this technique constitutes an improvement over existing approaches adopting DAOP. The strength points of COP adoption are analyzed and the advantages over DAOP techniques are discussed.

The paper is organized as follows. In Section 2 we introduce COP and its main features. In Section 3 we present the use of COP for the development of autonomic applications. Section 4 discusses the strength points of the COP approach. An extended case study is in Section 5. Section 6 analyzes the issues and the open problems associated to the use of COP in AC. Section 7 presents the related work. The conclusions and the future work are discussed in Section 8.

## 2. CONTEXT-ORIENTED PROGRAMMING

COP is a recent programming paradigm specifically aimed at supporting dynamic software adaptation to the execution context. COP constitutes an alternative to the use of special design patterns or to hard-coded conditional statements spread over the application to encode context-dependent behavior. In COP, *ad hoc* explicit language-level abstractions are introduced to express context-dependent behavioral variations and their run time activation. This approach makes the code easier to develop and maintain. Since behavioral adaptations are often scattered over the application code, they can be identified as crosscutting concerns. COP languages are specifically conceived to cope with this issue, and properly modularize the possible behavioral changes.

In COP, the definition of *context* is open and pragmatic: *any computationally accessible information* [Hirschfeld et al. 2008] is considered to be context. The identification of context changes can be used by the application to trigger adaptations. Hereafter we narrow down the concept of context and context changes to such things such as the values of environmental data provided by sensors or certain operational conditions internal to the application.

The essential concepts of the COP paradigm are summarized in the following points:

*Behavioral variations.* Variations express a *chunk* of behavior that can substitute or modify a portion of the basic behavior of the application.

*Dynamic variation activation.* Variations can be enabled at run time to affect the behavior of the system. This mechanism is the way COP programs react to context changes adapting their behavior. The operation that enables a variation is referred to as variation *activation*.

*Dynamic variation combination.* Variations that are active *at the same time* on the same behavioral aspect are combined with each other and with the basic behavior of the application. The final program logic results from this combination.

Despite different existing approaches, variations are usually introduced in COP languages through the concept of *layer*. Layers are language entities that group related context-dependent behavioral variations. Layers are first-class objects and can be explicitly referred in the underlying programming model.

In the following, we show an example of the ContextJ language [Appelauer et al. 2009; Hirschfeld et al. 2008], a contextual extension of Java, one of the most mature COP implementations. The example is extremely simplified to highlight the COP constructs (Figure 1). The code snippet defines a `Figure` class and a `Border` class. In the example, the notion of context is used to capture the variability in figure drawing. The `Figure` class declares the method `print`, which is redefined both inside the `bordered` layer and the `shadowed` layer. The `Border` class also declares a `print` method and redefines it in the `shadowed` layer. When `print` is called on an instance of the `Figure` class, the implementation to execute is chosen according to the currently active layers. Since the `with` statement activates a given layer over the scoped block, when `print` is called on the `border` object the `shadowed` layer and `bordered` layer are active (despite the `bordered` layer has no effect on the `border` object). Methods are searched inside layers in reverse activation order.

The activation of the `bordered` layer adds a border to the figure to print while the `shadowed` layer adds a shadow to the printed figure. Of course also a single layer activation is possible. The call to `proceed` is similar to `super` in Java and executes the method implementation in the next active layer. If there are no further active layers, the original method is called. Layer activation is dynamically scoped: it affects the behavior of the program not only for the method calls syntactically inside the code block, but also for all the calls triggered in turn.

In addition to the constructs exemplified in Figure 1, ContextJ supports a `without(layer){ ... }` statement that temporarily disables the given layer in the associated code block. The replication of the `with` and `without` statements for multiple layer (de)activations, can be avoided by directly specifying a list of layers inside the round brackets. Since layers are first class citizens, they can be passed as function parameters and returned as values, and assigned to variables. This feature is especially required to bind the layers that need to be activated to the result of a computation, allowing run-time planned adaptability.

In addition to *around* methods – like the ones in the example, which are called instead of the original method – *before* and *after* methods can also be defined similarly to Lisp standard method combination or to AspectJ advices. ContextJ provides a reflective API, which gives access to layer inspection and manipulation. For example it is possible to know the active layers at a certain point of the execution or to activate a layer obtained from introspection, or query a layer for the partial method definitions it declares.

Over the years several COP extensions to various languages have been proposed. Each of these implementations interprets the COP paradigm according to the underlying programming model and provides language-specific functionalities. For this reason COP languages – though implementing the core concepts of the paradigm – come in a variety of different flavors. To give an idea of the functionalities offered by the paradigm beyond the example of the previous section, we present an overview of the most significant design choices: layer declaration strategy, the relationship between adaptation and concurrency and the extent of variation activation. A complete list of references will be provided in Section 7.

We distinguish between two layer declaration strategies: *layer-in-class* and *class-in-layer*. In the layer-in-class pattern a layer is declared inside the lexical scope of the module it affects. For example in ContextJ and JCOP [Appelauer et al. 2010], an extension of ContextJ, layers are declared in classes, the fundamental code unit in Java. In the class-in-layer pattern layers are defined outside the lexical scope of the module for which they provide behavioral variations. For example ContextL [Costanza and Hirschfeld 2005], an extension of Common Lisp, allows one to declare layers separately from classes, each layer syntactically embracing the variations related to different classes. Actually ContextL comprises both declaration strategies, allowing one to choose the separation degree between the basic behavior and the variation concerns.

Most COP languages employ a *per-thread* notion of context. For example in ContextL the `with` statement influences only the flow of control on which it is called. On the contrary in Ambience [González et al. 2007], another context-aware Lisp extension, the context—and therefore the corresponding active variations—is global and shared among all the threads. This design decision

```
public class Figure {

  layer bordered {
    public void print(){
      System.out.println("Figure: " +
                          "adding border");
      Border border = new Border(this);
      border.print();
      proceed();
    }
  }

  layer shadowed {
    public void printLayerShadowed(){
      proceed();
      System.out.println("Figure: " +
                          "applying shadow");
      ...
    }
  }

  public void print(){
    System.out.println("Figure: drawing");
    ...
  }
}

Figure f = new Figure();
f.print();
with(bordered, shadowed){
  f.print();
}
```

```
public class Border {

  private Figure figure;

  public Border(Figure figure) {
    this.figure = figure;
    ...
  }

  layer shadowed {
    public void print(){
      proceed();
      System.out.println("Border: " +
                          "applying shadow");
      ...
    }
  }

  public void print(){
    System.out.println("Border: drawing");
    ...
  }
}

—— Exec ——

Figure: drawing

Figure: adding border
Border: drawing
Border: applying shadow
Figure: drawing
Figure: applying shadow
```

Fig. 1.   An example of the ContextJ language.

reduces the granularity of the possible adaptation but simplifies the process of variation activation by a managing thread.

The vast majority of COP languages exploits a thread-based concurrency model. An exception is our ContextErlang language [Salvaneschi et al. 2012; Ghezzi et al. 2010b; 2010a] – a COP version of Erlang [Erlang ] – based on the concept of context-aware reactive agents, an implementation of the actor concurrency model [Hewitt et al. 1973]. Actors have a behavior which is executed when a message is received and messages are buffered in a mailbox. Actors communicate (only) through messages with no shared memory, which greatly simplifies the design of correct concurrent applications. In ContextErlang variations can be activated on the agents through context-related special messages. The behavior of the agents upon message reception results from the sequence of variations currently active on them.

In the example of Figure 1 variation activation has a dynamic extent because after the end of the code block the activated variations are automatically deactivated. Some COP languages, including ContextJ through the use of the reflection API, also support variation activation with indefinite extent: from the activation onwards the partial program definitions inside the variation affects the program until a different variation activation occurs.

Other features of the COP languages are discussed in Section 6 in the context of the support that they can provide for the development of autonomic systems.

### 3. CONTEXT-ORIENTED PROGRAMMING FOR AUTONOMIC SYSTEMS

The MAPE-K (Monitor, Analyze, Plan, Execute, Knowledge) loop is a reference model for autonomic control loop originally proposed by IBM [IBM 2003]. In the MAPE-K loop the managed element represents a software system to which is given autonomic behavior coupling with an autonomic manager [Huebscher and McCann 2008]. The autonomic manager is a software component that *monitors* the managed element, *analyzes* the collected data, *plans*—thanks to the internal *knowledge* of the system—the actions to take to satisfy the requirements, and *executes* the necessary changes on the managed element. *Sensors* collect information about the managed element, and *effectors* carry out the planned changes.

We propose a conceptual framework for the implementation of the MAPE-K loop. The framework leverages the COP paradigm to accomplish the adaptability requirements of the managed element. It includes the modularization of the behavioral variations that the autonomic manager can trigger on the managed element, and the implementation of the effectors through which this change is actuated.

The result of the planning phase of the autonomic element consists of a set of layers to be activated. The internal knowledge of the monitor collects all the layers that can be activated on the managed element and the possible constraints on them.

The managed element is implemented through the COP abstractions. When the control flow enters a *with* statement, the autonomic manager is queried for the active layers and the code in the scoped block is automatically adapted:

```
with ( AutonomicManager.getActiveLayers() ){
  // Dynamically adapted code ...
}
```

Each object involved in the computation triggered by the code block adapts itself depending on the layers declared in its class. With this model the autonomic manager directly decides the actual adaptation that should be performed, but it is up to the managed element when to actually grab this information and how to use it.

From a methodological standpoint the designer establishes which are the possible adaptations of the managed element and each adaptation is reified into a layer. The designer also identifies which parts of the application must be adaptive and have to be included inside the with(*activeLayers*){ ... } statements. All the adaptation concerns, both alternative behavior declarations and variations activation, appear explicitly in the code.

Our framework satisfies the guidelines for the architecture of an autonomic element, intended as the smallest unit of an autonomic application. An autonomic element is a self-contained module with specified context dependencies. We also assume that it embeds the mechanisms for self-management, and is responsible for implementing its functionalities managing its behavior in accordance with external context and policies [Parashar and Hariri 2005]. The hypothesis that all possible adaptations be known in advance is based on the rather strong assumption that all possible variations of the relevant environment states can be anticipated at design time. Under this assumption, the autonomic element stands in a *local loop* in which it can handle adaptation of behavior to recognized switches between known environment states. A *global loop*, which may involve machine learning, artificial intelligence, and/or human intervention is instead in charge of global optimization and can handle unknown environment states. Solutions for relaxing the *a-priori* knowledge of the possible adaptations in the autonomic element aren discussed later in Section 6.

Figure 3 shows an example of the use of COP in an autonomic system. In order to simplify the comparison with DAOP in the next section, we adopted the DAOP example in [Greenwood and Blair 2003] and implemented it with COP. Figure 3 shows the code of the `ResourceStorage` class, a simplified implementation of a software component that can be queried for a stored resource. If the response time exceeds a threshold value, a caching service is activated to improve the performance. The `ResourceStorage` class defines a `request` method for items retrieval. The method redefinition in the `MinimizeRespTime` layer consults the cache and, in case of a miss, calls the original implementation.

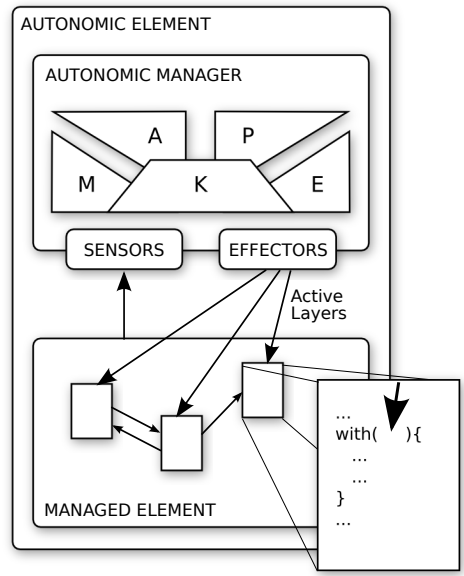

Fig. 2.  COP and the MAPE-K autonomic loop.

The way the autonomic manager decides which layer must be dynamically activated is related to the planning algorithm implemented in the monitor. For example event-condition-action (ECA) [Huebscher and McCann 2008] rules can be defined, which produce adaptation plans based on the monitored events. We do not provide further details on this topic, since investigating the design of the autonomic manager is out of the scope of the present paper.

## 4. DISCUSSION: CONTEXT-ORIENTED PROGRAMMING VS. DYNAMIC ASPECT-ORIENTED PROGRAMMING

In this section we discuss the strengths of COP in AC by unfolding the analysis together with a systematic comparison with aspect-oriented techniques. We provide preliminarily review of the context of application of the DAOP paradigm to the AC field.

The use of aspect-orientation in the development of autonomic systems can be traced back to two different goals: monitoring and dynamic adaptation. Monitoring embodies the implementation of the sensors of the MAPE-K model.

We now consider the cache example of the previous section. AOP can be used to collect information on the performance of the resource storage. This approach is shown by the code snippets in Figures 4 and 5, adapted from the work of Greenwood and Blair. An around advice is defined, which intercepts the calls to the `request` method of the resource storage and computes the time taken to retrieve an item. If this action takes too long, a caching facility is activated. In Greenwood and Blair's work the implementation of monitoring facilities is based on DAOP, arguably for uniformity reasons with the rest of their paper. However it worth noting that monitoring does not necessarily require dynamic activation of aspects and traditional AOP suffices. This consideration of course only holds under the assumption that the elements to probe are known in advance and run-time reconfiguration does not encompass which data are to be monitored.

Besides monitoring, aspects are used to change the behavior of the application dynamically according to the decisions taken by the autonomic manager. This is the way DAOP can be applied to the implementation of the effectors of the MAPE-K model. In this case, dynamic activation is a mandatory requirement, since it is essential to perform run-time behavior change. Adaptation is

```
public Class ResourceStorage{

  Cache cache = ...

  public String request(int req){
    Thread.sleep(time);
    switch (req){
      String item;
      // Retrieve item ...
      return item;
  }}

  layer MinimizeRespTime {

    public String request(int req){
      Object result = cache.get(req);
      if (result == null){
        result = proceed(req);
        cache.put(req,result);
      }
      return result;
  }}
}

...
// Can return the MinimizeRespTime layer
with (Manager.getActiveLayers()) {

  // Code that (even indirectly) uses
  // a ResourceStorage instance
  ...
}
```

Fig. 3.   Dynamically adding caching behavior using COP.

```
public String monitor(Interaction in) throws Error {
    long beforeTime = System.currentTimeMillis();
    String res = proceed(in);
    long afterTime = System.currentTimeMillis();
    long duration = afterTime − beforeTime;
    if (duration > threshold && !weaved){
      weaved = true;
      // Dynamically weave caching aspect
      ...
    }
    return res;
}
```

Fig. 4.   The monitor advice with DAOP.

obtained through activation and deactivation of advices. Figure 5 shows the caching advice activated by the monitoring code in Figure 4.

While aspect-oriented techniques were originally conceived for effective management of cross-cutting concerns, the success of DAOP in the AC field is substantially due to a link between software adaptation and dynamic aspects activation, rather than solving the issue of separate concerns. We believe that while the separation of concerns provided by the aspect-oriented paradigm perfectly fits the needs of application monitoring, COP can constitute a better solution for dynamic adaptation.

```
public String checkcache(Interaction in) throws Error {
  Integer arg = (Integer)in.args[0];
  String res = cache.get(arg);
  if (res == null){
    res = proceed(in);
    cache.put(arg,res);
  }
  return res;
}
```

Fig. 5.   The caching advice with DAOP.

## 4.1. Upfront Autonomic Design

DAOP makes it extremely easy to add functionalities to an existing system, since this can be done by leaving the original code base untouched. Thus DAOP appears as a natural solution whenever we wish to add autonomic capabilities to an existing system that was not conceived to support them. However, when we are building a system from scratch, we should instead support *upfront autonomic design*. This will allow us to write clearer code and to obtain a self-documenting application structure. This is the scenario in which COP comes into play. When the developer envisions the structure of an application, she should focus on (1) the parts of the application that need to be adaptable and (2) the adaptations each component must be able to perform autonomously. In COP both these points directly map on language constructs, without adding external machinery. Adaptable parts of the application are enclosed into *with* statements, while adaptations are defined inside layers, leading to self-explanatory code.

## 4.2. Encapsulation Enhancement

Since monitoring in most application can be considered as a separate concern, good engineering suggests to keep monitoring code apart from the functional logic of the application.

For adaptation code, the motivations for separation are less evident. For example, in the case of the resource storage (Figure 3), the caching behavior is conceptually separated from the basic behavior of the component, implemented in the `request` method. However, by keeping the variation in the same class of the basic code allows for immediate visualization of all the possible behaviors that the `ResourceStorage` class can adopt. Moreover, encapsulation of partial method definitions and original methods inside the same code unit helps avoiding potential inconsistencies that can be introduced by combining the behavior of around methods and the basic ones.

It is not infrequent that the distinction between basic behavior and variations simply does not hold. For example, suppose that the autonomic manager can pursue different goals: response time minimization or memory consumption minimization. The resource storage adopts different behaviors keeping the items in memory or on disk, according to the autonomic manager goals. In this case there is no distinction between basic behavior and variations: all possible behaviors are part of the functional logic of the application. Since there is no need for concern separation, the COP solution—which allows each code unit of the application to be defined together with the adaptation associated to that code unit—looks more appealing.

## 4.3. Control-flow-based Activation

The development of a managed element can require fine-grain adaptation. For example it can be necessary to activate different variations on a *per control flow* basis, affecting only one specific thread. Consider a server which assigns a thread to each connecting client. Clients running on mobile devices can have bandwidth limitations or bandwidth availability variable over time. Each client is served by a different thread which must take into account the specific bandwidth condition of the client. In this case, per control flow adaptation is a key feature to allow the server to be able to adapt to the situation of each client independently.

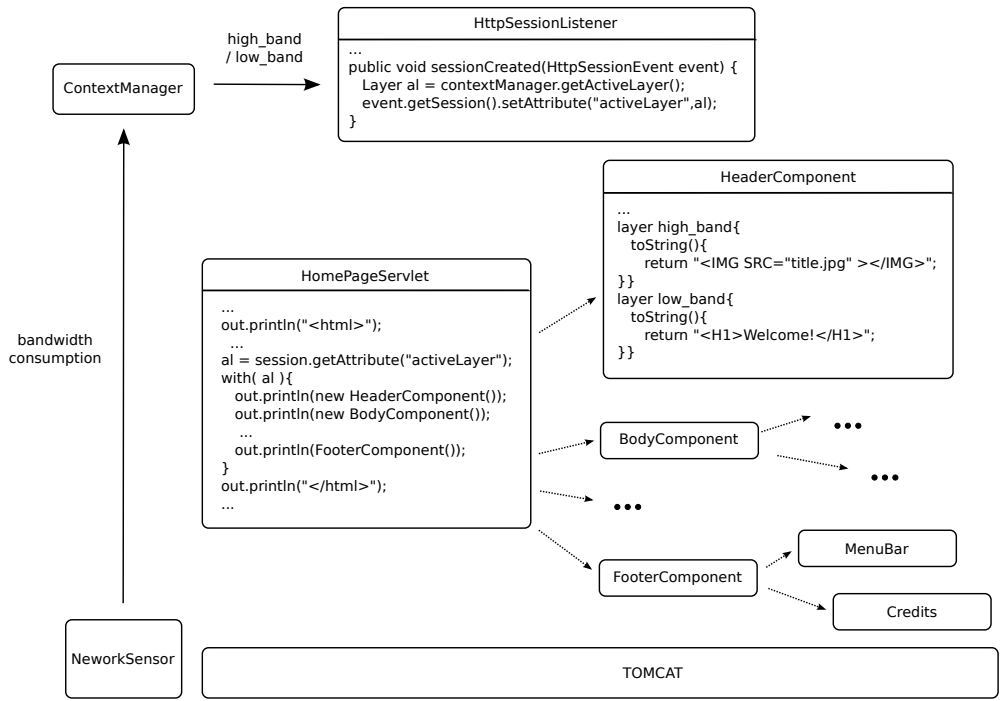

Fig. 6.   The structure of the adaptive Web application.

Control-flow-based activation has been investigated in many DAOP languages [Aracic et al. 2006; Truyen et al. 2001]. However only a subset of industrial-strength frameworks actually support it. For example the AspectJ language allows one to define `cflow` pointcuts, but this feature is currently missing in the Spring AOP pointcut model[1].

### 4.4. Dynamic Extent

The variations activated by the `with` statement have a dynamic extent in the sense that they obey a stack-based discipline. They are automatically removed when the scope of the statement is exited. This approach enforces a clean identification of the adaptable portions of an application. Conversely, indefinite activation such as the aspect activation statements adopted by many DAOP frameworks can easily lead to adaptation sequences that are difficult to foresee since the statements are scattered over the code and the ones actually executed depend on the flow of execution. The problem somehow reminds the use of *goto* statements compared against structured programming.

Another advantage of the dynamic extent is that it enforces the uniform activations of the variations over the whole scope of the code block, adapting uniformly *all* the entities involved in the scoped computation. This allows one to avoid the inconsistencies that can arise if an external thread (de)activates the variations asynchronously, in which case the adaptation extends over an unpredictable portion of the execution. The case study of the next section further clarifies this point.

Explicit dynamic extent for advice activation is supported by some DAOP frameworks. For example the CaesarJ [Aracic et al. 2006] language allows one to define a `deploy(asp) { ... }` statement which activates the `asp` aspect in the dynamic extent.

---

[1]The Spring Framework `http://www.springsource.org/`.

## 5. CASE STUDY

In this section we show how COP can be used to implement autonomic systems in practice. We describe an autonomic Web application and how it benefits from the COP concepts as discussed in Section 4. We evaluate the capability of the application to expose the autonomic behavior in a realistic scenario. We discuss in details the advantages of the COP solution compared with a pure Java and several AOP/DAOP implementation alternatives, then measure the performance of COP with respect to the other solutions. Finally we evaluate the impact of COP on the development process.

### 5.1. Autonomic Web Application

We have developed an adaptive Web application with the ContextJ language, using the Java Servlet technology in the Tomcat[2] application server. The application generates the pages of the Website with an overall visual quality that is chosen according to the current network bandwidth consumption. If the request rate increases and the load generated by the server on the network becomes excessive, lighter pages are served, for example by substituting flash animations with static images, or high-quality pictures with simple graphical elements.

The scenario for which we conceived this application is a datacenter with fixed bandwidth availability. The servers in the datacenter have an assigned bandwidth threshold that should not be exceeded. Each server constitutes one of the autonomic elements of which the whole datacenter – the autonomic system – is composed. Beside the *local loop* implemented by each autonomic element, it is reasonable to envisage the implementation of a *global loop* that helps to accomplish higher level goals: for example, activating new servers, negotiating an increase of the overall bandwidth limit, or reassigning the threshold of each server when needed.

The autonomic manager is implemented as a `ContextManager` thread (Figure 6), which periodically plans the layers to activate on the application. To make its decision, the context manager relies on a `NetworkSensor` component, which monitors the current network bandwidth consumption. The context manager conceptually operates on the basis of a simple ECA rule, which switches from the `high_band` layer to the `low_band` layer when a threshold is reached. For a more effective result, we implemented in the `ContextManager` manually-tuned Proportional Integral feedback controller that monitors the current bandwidth usage and controls the fraction of newly created `low_band` and `high_band` sessions, according to the error from the defined setpoint.

Each page is organized in page components, which individually encapsulate the adaptation capabilities. For example, a page can be composed by a `HeaderComponent`, a `BodyComponent`, a `FooterComponent` and a `SideComponent`. Each page of the Website is associated with a Java Servlet, which inside a `with` block requests each page component to print itself and generate the associated HTML code.

### 5.2. COP Support and Design Choices

COP supports *upfront autonomic design* since the possible adaptations map directly onto layers, making the architecture clean and intuitive.

Because of *encapsulation enhancement*, each component of the page directly includes the adaptations it is able to perform. Any subsequent modification of the way an element is rendered in a given contextual condition requires only a change that is local to the component.

*Control-flow-based activation* is essential in this application, since each servlet is executed in a different thread for each connected client, and each thread must be adapted independently from the others. Otherwise, after exceeding the bandwidth threshold, the only possible adaptation would be that *all* the requests are served by servlets in `low_band` mode, which results in an abrupt breakdown of the bandwidth consumption and the possible emerging of oscillatory phenomena.

*Dynamic extent* allows uniform adaptation to be enforced across the elements of the same page. We briefly discuss this, since the choice of granularity of the adaptation was an issue that arose during the design. We identified tree alternatives: adaptation at component level, at page level, or at session level. Adaptation at component level implies that each page component independently queries the context manager for the active layers before generating its associated HTML code. This leaves open the possibility that in the same page some components appear in `low_band` mode, and others in `high_band` mode, which may result in a poor visual effect. Adaptation at page level queries the context manager once for each page generation and therefore constrains all the page components to be drawn with the same active layer. This solution is reasonable except that, if a user hits the *back* browser button, he could be shown a page looking different from the one he has just seen, a fact that could be disorienting. For these reasons we chose adaptation at session level: when a new session is created, the context manager is queried for the active layers, which are then stored in the session object. Since each servlet retrieves the active layers form the session, we achieve enforcement of coherence of all the pages visited by the same user.

---

[2]The Tomcat application server `http://tomcat.apache.org/`.

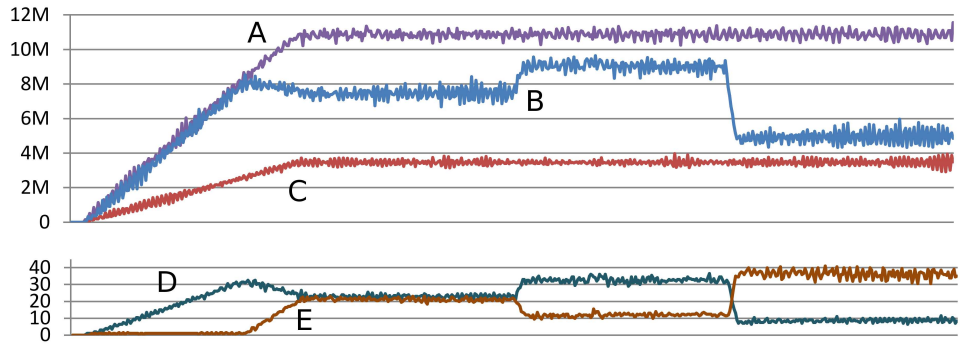

Fig. 7.   Experimental valuation of the Web application. Bandwidth consumption (top) and proportion between new `high_band` and `low_band` sessions (bottom).

The drawback of this solution is that the adaptation process is potentially slow, because a session lasts for all the time the user browses the site, and pages continue to be served in the same way.

### 5.3. Experimental Evaluation

We evaluated our approach simulating the activity of 200 simultaneous users with an HTTP traffic generator. Their number was gradually increased by adding a new user every second, for a ramp-up interval of 200 s. Simulations last for about 800 s (Figure 7). Each user requests 5 pages from the site, with a delay of 1 s between two consecutive requests. A further request makes the session expire, and the user activity restarts.

To show the impact of the autonomic behavior we compare three different runs (Figure 7 (top)). Two concern non-autonomic versions of the application operating only in `high_band` mode (A) and in `low_band` mode (C). The area between lines A and C is the controllable region of the system. Run B shows the effect of the controller. The initial setpoint is 7.5 MB/s raised to 9 MB/s after 400 s and lowered to 5 MB/s after further 200 s. Figure 7 (bottom) shows the number of `high_band` mode sessions (D) and `low_band` mode sessions (E) per second generated during the run with the controller. As expected, after the ramp up phase, the two lines sum to 40 sessions/s. The traffic was generated from a laptop connected through a 100 Mb LAN to a PC equipped with an Intel Core 2 Quad Q9550, 4 GB RAM. The Web application runs on Tomcat 6.0 and Java 6 over Windows XP SP2.

Although the control problem was not our primary goal, the simple solution we adopted showed to be effective and allowed us to evaluate the use of COP in a concrete and fully working autonomic application.

```
public class Home {
  public int activeVariation = LOW_BAND;
  private PrintStreamStub out = new PrintStreamStub();
  public static final int LOW_BAND = 0;
  public static final int HIGH_BAND = 1;

  public void doGet(){

    out.println("<html>");
    out.println("<head>");
    out.println("<title>" + "Autonomic Computing Inc" + "</title>");
    out.println("</head>");

    out.println("<body>");
    Component1 p1 = new Component1();
    out.println(p1);
    Component1 p2 = new Component2();
       ...
    out.println("</body>");

    out.println("</html>");
} }
```

Fig. 8.   Stub implementation of the `Home` servlet.

## 5.4. Option Analysis

In this section we compare the COP implementation of the autonomic Web application with possible alternatives. After the more abstract analysis of Section 4, the intent of this section is to provide a practical discussion of the concrete choices available to the developer. For the AOP/DAOP options we chose AspectJ, since it is an industrial-strength AOP framework with support for dynamic capabilities. To make the discussion clearer we consider a generic model which captures the structure of the application, abstracting the details of each single Web page. We instantiated this model as several stub implementations leveraging different AOP/DAOP features.

The `Home` class models the servlet. The implementations of the `Home` class have some differences which are discussed hereafter. For example the `activeVariation` variable in some cases needs to be thread-local. In Figure 8, as a reference we show the implementation used in the *wormhole* pattern case (described later). Inside the `doGet` method of the `Home` class the first-level page components `ComponentN` are instantiated and the `toString` method is called. Inside the `toString` method of first-level components, second-level components `ComponentNM` are instantiated and the `toString` method is called. We generically refer to fist-level and second-level components as `Component` classes. The `Home` objects request the current bandwidth level to the context manager and store it for future use.

*Pure Java.* The first solution we analyze is a naive implementation in pure Java, without recurring to aspect orientation (Figure 9). This solution allows us to uncover some design flaws which are gradually addressed by the next implementations. The alternative behaviors of each `Component` object in low and high bandwidth conditions are implemented in the `toStringLow` and in the `toStringHigh` methods. The `toString` method implements the dispatching logic to select the proper behavior.

This solution presents some inconveniences. The dispatching algorithm is made explicit, and the `toString` method is split into the `toStringLow` and the `toStringHigh` methods. This exposes to the client the autonomic logic and unnecessarily complicates the structure of the class. Another issue deals with the propagation of the autonomic information. The `Home.BANDWIDTH` variable must be made accessible by the `Home` class by marking it as static, so practically a globally-visible variable. This also requires to set it thread-local to avoid interference between multiple servlets running in different threads. A possible alternative could be to pass the bandwidth level as an additional pa-

```
public class Panel1 implements PageComponent{

  public String toString(){
    if(Home.activeVariation.get() == Home.HIGH_BAND){ return toStringHigh(); }
    else{ return toStringLow(); }
  }
  public String toStringLow(){
    return "component1-low_band </br>" +
      new Component11().toString() +
      new Component12().toString() ;
  }
  public String toStringHigh(){
    ...
} }
```

Fig. 9.   Stub implementation of the autonomic application in pure Java.

```
public aspect InterceptorAspect {

  pointcut printing(PageComponent pc) :
      execution( public String PageComponent.toString() ) && this(pc);

  String around(PageComponent pc) : printing(pc) {
    if(Home.activeVariation.get() == Home.HIGH_BAND){
      return pc.toStringHigh();
    } else {
      return pc.toStringLow();
    }
} }
```

Fig. 10.   Stub implementation of the autonomic application separating in the aspect the dispatching logic.

```
public aspect InterceptorAspect {

  pointcut highState() : if ( Home.activeVariation.get() == Home.HIGH_BAND );

  String around() : execution( public String Component1.toString() ) && highState() {
    return "component1-high_band </br>" +
      new Component11().toString() +
      new Component12().toString() ;
} }
```

Fig. 11.   Stub implementation of the autonomic application using IF AspectJ pointcuts.

rameter in method calls. However this pollutes the interfaces of the `toString` methods and burdens
first-level `Component` objects with the responsibility of taking care of the propagation of this value.

*AOP - Separating the Dispatching Logic.* This solution is quite similar to the previous one, except
that the dispatching logic is separated from the main application using an *advice*, which intercepts
the call to the `toString` method and redirects it accordingly (Figure 10). Since the dispatching
algorithm is hidden in the advice, the code in the `Component` class is cleaner. However, the problem
of autonomic information propagation still remains. Interestingly, this solution, like the previous
one, preserves the symmetry between the `toStringHigh` and the `toStringLow` methods, which is
sacrificed by other design choices (see next section).

*Dynamic Pointcuts - IF Pointcut.* In AspectJ, the `IF` pointcut triggers advice activation depend-
ing on a dynamic condition. Figure 11 shows how this type of advice can be used to dynamically
activate an alternative implementation of the `toString` method when the servlet is assigned the

```
public aspect InterceptorAspect {

  pointcut caller(Home home) : execution( public void Home.doGet()) && this(home);

  pointcut wormhole(Home home) :
      cflow(caller(home)) && if(home.activeVariation == Home.HIGH_BAND);

  String around(Home home) :
      wormhole(home) &&
      execution( public String Component1.toString() ) {
    return "component1-high_band </br>" +
      new Component11().toString() +
      new Component12().toString();
} }
```

Fig. 12.   Stub implementation of the autonomic application using the `CFLOW` AspectJ pointcut and the wormhole pattern.

`HIGH_BAND` condition. This solution allows one to solve the problem of declaring the dispatching algorithm explicitly. The drawback is that it introduces an asymmetry between conceptually alternative behaviors. In fact, the default behavior is associated to the low bandwidth condition, while the high bandwidth condition behavior is encapsulated in the advice.

*Dynamic Pointcuts - CFLOW Pointcut - The Wormhole Pattern.* This solution adopts the AspectJ `CFLOW` pointcut, which quantifies over the joinpoints in the control flow of a given pointcut (Figure 12). The proposed solution is an instance of the *wormhole pattern* [Laddad 2009], whose purpose is to propagate a value (in this case the autonomic information) without polluting interfaces. The `wormhole` pointcut binds a `Home` instance. When a `toString` method is called in the control flow of the `doGet` method of the instance, it is possible to access the autonomic information to decide if the advice must be activated. Interestingly the per control flow activation allows to avoid to declare the `activeVariation` field thread-local.

This solution shows how aspects can be used to cleanly propagate the autonomic information without polluting interfaces nor using global variables. However, as already discussed, the symmetry between the alternative behaviors is lost.

*Context-oriented Programming.* We briefly discuss how the COP solution we adopted for the autonomic application (Figure 6) solves the issues emerged in the previous approaches. The dispatching logic is automatically managed by the layer mechanism and does not appear explicitly in the client code. The autonomic information propagation is transparent to the developer, thanks to the dynamic scope layer activation mechanism. This allows one to avoid to explicitly define thread-local or global variables. Finally, since several method definitions can be provided assigning them to different layers, symmetry is preserved.

## 5.5. Performance Evaluation

In this section we evaluate the performance impact of COP in the case study. A possible approach is to implement the autonomic application using different solutions – such as those discussed in the previous section – and compare their performance through network traffic simulation. However our early experiments suggest that the network communication and the Web server runtime introduce an overhead and a variability which hide the impact of COP. This is an encouraging preliminary finding, showing that in a real setting the overhead can be negligible.

To further investigate this point we proceeded as follows. We used the stub implementations described in the previous section as a benchmark for the portion of code which deals with page creation, the *core* functionality of the autonomic application. This concentrates the analysis on the part of the application which is directly affected by COP, neglecting the overhead of the Web server framework, of the servlet creation, and of the network communication.

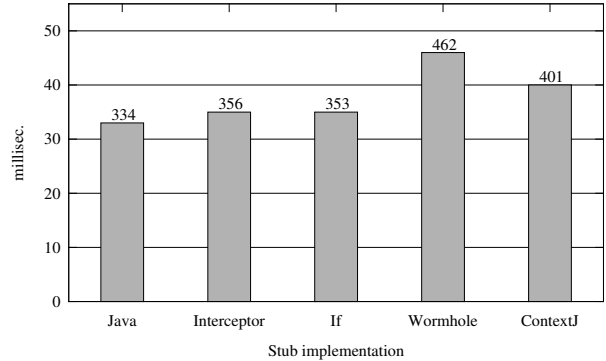

Fig. 13. Performance comparison for the solutions in the option analysis.

In the benchmark, we instantiate the `Home` class (i.e. Figure 8) and we call the `doGet` method on it. This creates four first-level components, each of them creating two second-level components. The HTML code generated by the `Home` page and by the components is simply discarded by the `PrintStreamStub` object. We repeated this process $10^5$ times and considering the mean over 10 executions.

It is worth noting that such test greatly amplifies the overhead introduced by non plain-Java techniques, since most operations not affected by the slowdown, are removed from the benchmark. Nevertheless, the results in Figure 13 show that the performance penalties of COP are quite contained. Not surprisingly, the pure Java implementation is the fastest. COP is outperformed by static AOP and by the implementation based on the `IF` dynamic pointcut. However those solutions comes at the cost of the design shortcomings discussed in the previous section. The wormhole patter, which is probably the best DAOP solution from a design perspective, is outperformed by COP.

*Microbenchmark.* The previous analysis evaluates the overall effect of using COP in our case study. To further inspect the performance impact of COP we implemented a microbenchmark focusing on method dispatching, which is the feature actually affected by the COP overhead.

In the benchmark we call a contextually dispatched method for which five partial definitions exist in different layers. All the partial definitions proceed. We evaluated the time spend by $10^7$ calls to the method varying the number of active layers. This result is compared with a similar setup in plain Java. Six methods call the next one up to the base one. We simulate a variable number of layers by changing the method initially called.

In Figure 14 we show the results varying the number of active variations from 1 to 5. Interestingly, even considering only method dispatching, the slowdown of COP can be evaluated in about 1/3 with respect to pure Java.

All the performance evaluations described in these sections were executed on a machine used for the experimental evaluation (Section 5.3) with AspectJ 5 installed. To allow the JVM to perform optimizations, each measure was taken with a dry run before the real test.

### 5.6. Impact of COP on the Development Process

In this section we evaluate the impact of the use of ContextJ in the development process, compared to pure Java. ContextJ is a source-to-source compiler. It implements the COP constructs mapping them to standard Java code and then transforming to bytecode through a standard Java compiler. A folder is automatically generated which contains the additional classes used for COP. In the case of the autonomic Web application this simply required to deploy this folder into the applications server together with the other `.class` files which compose the application.

The development of a software artifact in a real-world environment is a complex task which requires an echosystem of tools. Beside compilers, programmers use IDEs, and other tools which support the development process, such as debuggers, testing suites, code coverage analyzers or performance profilers. These tools are aware of the language syntax and semantics and a language extensions can make them completely unusable.

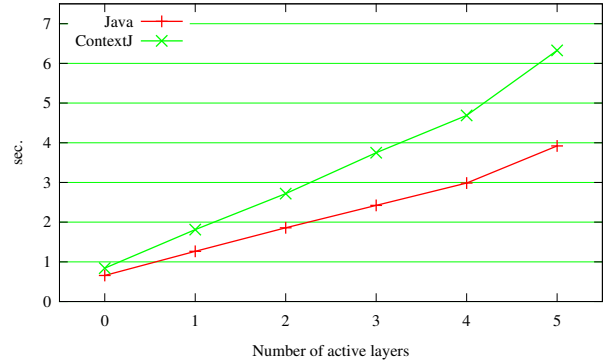

Fig. 14. Microbenchmark: performance of method dispatching in COP.

In the development of the autonomic Web application we encountered this issue in the use of IDEs, which do not recognize the COP extensions. Therefore we had to use a standard text editor, giving up syntax highlighting (which however can be easily implemented) and advanced features such as code completion or early error warning.

Tools such as a debugger or a performance profiler demonstrated to be still usable, but the analysis is cluttered by the additional operations added by the source-to-source compiler.

The considerations made so far are real issues and should be taken into account when developing a COP-based autonomic application. However such issues are not due to intrinsic technical limitations, but to the current status of the COP implementations. Since the landscape of COP is fastly evolving, it can be reasonable to expect that in the future increasingly better tool support for COP languages will be provided. Some steps in this direction have already been done: for example the EventCJ [Kamina et al. 2010; 2011] and the JavaCtx [Salvaneschi et al. 2011] languages are currently released as plugins of the Eclipse IDE.

## 6. KNOWN ISSUES AND OPEN PROBLEMS

In this section we highlight some of the issues that can arise using COP in implementing autonomic systems. When such issues have been already investigated by the COP community, we indicate the language extensions that were devised to cope with them. We anticipate that no currently existing language extension supports all the features one would like to use for designing dynamically adaptable software. We will discuss the problem and we will sketch a research direction for the solution.

*Variations Constraints.* In certain applications it can be necessary to express constraints on layers. Two layers can conflict with each other and must not be activated together or the presence of a layer can be required for the activation of another one. For example, the layer which reduces the response time of the resource storage keeping the items in memory is incompatible with the layer that uses the disk in order to reduce memory consumption. In other cases, a dependency between layers can exist, so that the (de)activation of a layer requires the (de)activation of another one.

This issues were initially solved by the COP community through the introduction of computational reflection for COP languages, which allows one to express layer constraints and to control their activation during the application execution [Costanza and Hirschfeld 2007]. Despite its extreme flexibility, this approach is quite complex to master. For this reason, declarative constraints on layers have been proposed as an extension to ContextL [Costanza and D'Hondt 2008] to make layer dependency enforcement easier. In this solution the violation of a constraint raises an error signalled to the programmer, who is expected to interactively fulfil the unmet constraint. Thanks to the resumable nature of Lisp, the execution is subsequently continued. Another approach consists in the use of a Domain Specific Language developed especially for the purpose of expressing layer dependencies [González et al. 2010]. This solution has the advantage of being more intuitive and less verbose than programmatic constraint declaration.

We believe that future research should be performed by taking into account the achievements of AC in the field of planning, especially through ECA rules, that could be used as a first approach to encode layer constraints. Another open issue is the reaction to a constraint violation. Since self-healing behavior is one of the key points of AC, it is obvious that solutions must be investigated that do not require human intervention.

*Unforeseen Adaptation.* In Section 4.1 we made the (restrictive) assumption that the developer knows all the possible software adaptations in advance and designs the application accordingly. In case this hypothesis does not hold, it is required to dynamically upload the code units that implement the initially unforeseen behaviors, following the general pattern provided by Java reflective facilities for dynamic class loading.

Some COP languages provide specific support for addressing this issue. For example, ContextErlang leverages the dynamic code loading capabilities of the Erlang platform, to implement remote transmission of the modules which implement the variations. With this feature, an agent on a remote Erlang node can be provided with a new behavior just by sending the variation and then activating it.

*Event Driven Adaptation.* Layer activation in accordance to the program control flow (i.e. using the `with` statement) is often insufficient to express the needed adaptations. Context changes are often triggered from external events. This has been acknowledged by existing proposals to augment COP languages with event-driven asynchronous context change primitives.

The JCOP language [Appeltauer et al. 2010], an extension of ContextJ, faces the problem with ad hoc language constructs for *conditional composition*, which declaratively expresses global variation activation on the basis of an enabling condition. With this feature the developer is relieved from specifying variation activation programmatically from inside the code. A similar mechanism is available in EventCJ, another Java-based COP language, which allows context to switch when an event is received. In EventCJ, it is possible to declare context transition rules: when an event is fired, a group of layers is automatically (de)activated on each object of the application. The solution of ContextErlang is based on the integration of the actor concurrency model with the COP paradigm. Since ContextErlang agents can exchange special context-change messages which trigger adaptation on the other agents, a context provider component can *push* the variation activation on the context-aware agents.

It is worth observing that the flexibility of asynchronous event-driven adaptation comes at the cost of losing control on consistency. In fact, layer activation with dynamic scope allows enforcing all the entities involved in the computation inside the lexical scope of layer activation to be uniformly adapted to the active layers. On the other hand an adaptation event could happen unexpectedly in the middle of a computation, which therefore can involve both adapted and non adapted entities without direct control.

*State Adaptation.* COP focuses mainly on functional adaptation through context-aware method dispatching. However sometimes, beside contextual behavior, contextual state must be taken into account. In PyContext [von Löwis et al. 2007] one can define *context variables*, which maintain their state during the dynamic extent of the `with` statement. Rather than having a fixed binding to a value, context variables have a binding to a value that depends on the execution context: the dynamic extent of the `with` statement determines the actual value of the variable.

This approach makes sense only if the state changes during the execution, and it must be preserved across a layer (de)activation. For example in the case of the case example Web application this feature is not required despite the fact that the HTML representation of each page component is context-dependent. This representation is stored in a variable local to a method of a certain layer and does not need to be modified.

*Aspects of Layers.* COP and DAOP can be considered as complementary techniques (Section 4). COP is in fact best employed in modularizing different behavioral alternatives while DAOP is best employed in modularizing separate concerns, such as monitoring. Thus it may be useful to take into account the use of both paradigms in the same autonomic application. For example a monitoring aspect could be added to a layered method.

Since some COP implementations rely on aspect-oriented compilers, the coexistence of both COP and AOP in the same language should not raise specific technical challenges. However, to the best of our knowledge, this integration has not been investigated so far.

*Layer Behavior Specification.* A precise description of the behavior implemented by a layer is one of the open problems in the COP research field. While the availability of a formal description of a software artifact is a desirable property in general, it becomes necessary for the application of COP to the implementation of autonomic systems. In fact, since the autonomic manager is in charge of activating layers on the managed element, a formal specification of the variation introduced in the system behavior is required in order to plan the proper layer activation in a fully automated manner. A reasonable approach could be based on the introduction of metadata associated with each layer the autonomic manager can rely on to make the planning decisions.

## 7. RELATED WORK

The issue of dynamic software adaptation has been extensively addressed from an architectural perspective [Garlan et al. 2004; Oreizy et al. 1998; 2008]. McKinley et al. [McKinley et al. 2004] review the techniques applied to adaptive software composition. Separation of concerns, computational reflection and component-based design are identified as the key enabling technologies. Dowling et al. [Dowling et al. 2000] compare different language-level techniques for supporting software dynamic adaptation: reflection, dynamic link libraries (DLL), and design patterns. Although this work dates back to before the spreading of AOP and AC, it still stands as an interesting study in the field. The conclusion of the authors is that computational reflection introduces a non-negligible overhead, but nevertheless it offers significant advantages in separating functional and adaptation code.

Other researchers also explored design patterns as a viable solution to software adaptation. Traditional GoF patterns [Gamma et al. 2000] were extended to support context-awareness [Riva et al. 2006] and also totally new patterns were proposed [Rossi et al. 2005]. A broad perspective in the study of design patterns in autonomic systems was mainly carried on by Ramirez and Cheng [Ramirez and Cheng 2010], Their inspected over thirty adaptation-related research papers and project implementations from which they finally harvested twelve design patterns for autonomic and self-adapting software.

Dynamic aspect activation has been extensively investigated by researchers. Among the most significant works we mention CaesarJ [Aracic et al. 2006], Lasagne [Truyen et al. 2001] Prose [Popovici et al. 2003], JAC [Pawlak et al. 2001] and AspectWerkz [Bon 2004]. To some extent DAOP is now available also in industrial-strength tools such as AspectJ [Kiczales et al. 2001]. As already mentioned, DAOP has been introduced in the field of autonomic systems by Greenwood and Blair [Greenwood and Blair 2003], mostly as a mean to add autonomic behavior to an existing system. TOSKANA [Engel and Freisleben 2005] is a toolkit that applies the concepts of AC to the operating systems. TOSKANA uses DAOP for deploying dynamic aspects into the kernel and modify the its behavior while the system is running. J-EARS [Bachara et al. 2010] is a DAOP-based framework for autonomic systems development and management that allows sensors and effectors to be dynamically added and removed from a working Java application.

Software Product Lines (SPL) are families of software products sharing common behavior and differentiating in base of functionalities called *features*. SPL engineering has the goal of reducing time and effort in the development of applications in the same family. Traditional techniques use Feature-oriented Programming (FOP) to reason about features combinations and representing fea-tures in the language [Batory 2004]. The difference between FOP and COP is that FOP features are statically combined in the compilation phase, while COP allows runtime adaptation through dynamic combination of layers. Dynamic Software Product Lines (DSPL) [Cetina et al. 2010] have been recently explored to support adaptive systems by switching among the available features at run time [Bencomo et al. 2008; Hallsteinsen et al. 2006]. However, the approaches proposed so far are at the architectural level, while COP specifically focus on language-level abstractions.

JSpoon [Konstantinou and Yemini 2003] is an extension of the Java language and a runtime environment. In JSpoon it is possible to declare special variables, which are used to monitor and control the program behavior. These variables are automatically exported through the JSpoon runtime environment to an external management process which can modify their value changing the program execution. TRAP/J [Sadjadi et al. 2004] is a toolkit that allows one to add adaptive behavior to an existing Java application without modification of the original source code. The employed technique leverages behavioral reflection and AOP. AOP is used to organize the instantiation of wrapper classes instead the original ones, while reflection, in the form of a meta-object protocol [Kiczales et al. 1991], is used to inspect the application at run time and dynamically redirect methods calls to delegate objects. TRAP/C++ [Fleming et al. 2005] is a C++ implementation of the same concepts which uses generative programming to overcome the lack of reflection in C++.

Context-oriented programming was proposed in the pioneering work on ContextL [Costanza and Hirschfeld 2005; Costanza 2008; Costanza and Hirschfeld 2007] based on the CLOS meta-object protocol. Over the time, many COP extensions have been developed for different languages such as Python, Smalltalk, Ruby, JavaScript, Scheme, Groovy and others. This effort have extended to less dynamic languages, in which COP extensions are more difficult to implement due to limited reflective capabilities, such as Java [Hirschfeld et al. 2008; Appeltauer et al. 2009]. A comparison of the existing languages with a performance evaluation of the available solutions can be found in [Appeltauer et al. 2009].

Some COP languages deviate from the layer-base model. The Ambience Object System [Gonsáles et al. 2006; González et al. 2007] is built on top of Common Lisp. It is based on multi-methods dispatching, delegation and context objects. As already mentioned, ContextErlang is a COP language based on the actor model [Salvaneschi et al. 2012; Ghezzi et al. 2010a; 2010b].

While maintaining layers as the key enabling abstraction for context-awareness, recent COP research concentrated on alternatives to the `with`-based activation. Appeltauer *et al.* [Appeltauer et al. 2010] proposed pointcut-like expressions to activate layers. Kamina *et al.* [Kamina et al. 2010; 2011] employ a similar mechanism, but pointcut-like expressions trigger layer transitions *on single objects*, an therefore they abandon the per control flow model. Lincke *et al.* [Lincke et al. 2010] propose an open implementation [Kiczales 1996] in JavaScript, which provides a basic API the developer can use to experiment new activation mechanisms.

Our JavaCtx [Salvaneschi et al. 2011] is specifically aimed at simplifying the adoption of COP and its integration in the tools echosystem. Instead of extending the language with new constructs and break tool compatibility, JavaCtx expresses COP abstraction in plain Java, using coding conventions. An aspect library allows to inject in the program the semantics modifications for context-awareness.

While this is the first work which explicitly positions the COP concepts into autonomic computing and into the MAPE-K framework, several COP applications have been proposed which exploit COP to achieve autonomous behavior. Many examples are in the field of software for mobile devices [Kamina et al. 2010; 2011; González et al. 2007; González et al. 2010]. Typical adaptations consist in the dynamic switch between different information sources such as the Wifi connection, the GPS, or a local database. Also desktop software has been given adaptation capabilities using COP. For example, CJedit is a text editor which can dynamically adapt to the user activity (documenting or coding) [Appeltauer et al. 2010; Kamina et al. 2011]. Other examples encompasses smart home environments [González et al. 2010], graphical frameworks [Lincke et al. 2010], and software transactional memories [Costanza et al. 2009].

## 8. CONCLUSION AND FUTURE WORK

In this paper we proposed COP as a programming paradigm for autonomic systems. We presented a conceptual framework that positions COP in the autonomic MAPE-K model for the development of the adaptability concerns in the managed element. The advantages of this technique over the widespread DAOP paradigm were extensively discussed.

Our plans for the future unfold along two directions. On the one hand we will continue in the development of autonomic system prototypes using the COP paradigm in order to provide a set of significant case studies to the community. We believe that the existence of an extensive knowledge base of empirical studies that highlight the issues encountered in the development of autonomic systems and the possible solutions can significantly increase the level of engineering in this field. On the other hand we plan to investigate how the facilities offered by the COP model can be further specialized to meet the needs of autonomic applications. For example we plan to investigate a possible compromise between the coherence imposed by dynamic extent variation activation and the flexibility offered by event-driven indefinite activation.